\documentclass[conference]{IEEEtran}

\usepackage{cite}
\usepackage{amsmath,amssymb,amsfonts,amsthm}

\usepackage{graphicx}
\usepackage{mathtools}
\usepackage{numprint}
\usepackage{textcomp}
\usepackage{url}
\usepackage{xcolor}
\usepackage{xfrac}
\usepackage[sort&compress,numbers]{natbib}
\usepackage{algorithm}
\usepackage{algpseudocode}
\usepackage{blindtext}
\newtheorem{remark}{Remark}
\usepackage{tikz}
\usepackage{amssymb}
\usepackage{subcaption}
\usepackage{multicol}
\usepackage{subcaption}

\ifCLASSINFOpdf
\else
\fi
\begin{document}

%\title{Performance Analysis of Universal Decoders in the Presence of Alpha-Stable Noise Channels}
\title{Joint Error Correction and Fading Channel Estimation Enhancement Leveraging GRAND}

\author{\IEEEauthorblockN{Charles Wiame}
\IEEEauthorblockA{Research Laboratory of Electronics \\
Massachusetts Institute of Technology\\
wiame@mit.edu}
\and
\IEEEauthorblockN{Ken R. Duffy}
\IEEEauthorblockA{College of Engineering and College of Science\\
Northeastern University\\
k.duffy@northeastern.edu}
\and
\IEEEauthorblockN{Muriel Médard}
\IEEEauthorblockA{Research Laboratory of Electronics\\
Massachusetts Institute of Technology\\
medard@mit.edu}}


%Telephone: (800) 555--1212, Fax: (888) 555--1212}


\maketitle

\begin{abstract}
We present a novel method for error correction in the presence of fading channel estimation errors (CEE). When such errors are significant, considerable performance losses can be observed if the wireless transceiver is not adapted. Instead of refining the estimate by increasing the pilot sequence length or improving the estimation algorithm, we propose two new approaches based on Guessing Random Additive Noise Decoding (GRAND) decoders. The first method involves testing multiple candidates for the channel estimate located in the complex neighborhood around the original pilot-based estimate. All these candidates are employed in parallel to compute log-likelihood ratios (LLR). These LLRs are used as soft input to Ordered Reliability Bits GRAND (ORBGRAND). Posterior likelihood formulas associated with ORBGRAND are then computed to determine which channel candidate leads to the most probable codeword. The second method is a refined version of the first approach accounting for the presence of residual CEE in the LLR computation. The performance of these two techniques is evaluated for [128,112] 5G NR CA-Polar and CRC codes. For the considered settings, block error rate (BLER) gains of several dBs are observed compared to cases where CEE is ignored.
\end{abstract}


\IEEEpeerreviewmaketitle

\section{Introduction}
In wireless communication systems, information signals can experience several impairments: small-scale fading, interference, and additive noise generated by the transceiver electronics \cite{goldsmith}. To address the first of these disturbances, current mobile systems use pilot sequences prior to data transmission \cite{1359139}. These sequences, known in advance by the receiver, enable the estimation of small-scale fading channels. Once the channel is estimated, its effects on received data can be mitigated with equalizers \cite{5635468}. Longer pilot sequences can yield more accurate channel estimates, which in turn reduce error rates. However, the use of pilots reduces the effective data rate, as these sequences do not carry information.

This work addresses this trade-off by introducing a novel framework based on GRAND decoders. This framework enables effective operation with less precise fading estimates or equivalently higher channel estimation errors (CEE), while maintaining the same block error rate (BLER). This approach is particularly promising as it has the potential to improve data rates with minimal overhead, achieved by running GRAND multiple times. Thanks to GRAND's high parallelizability in digital implementations, this aspect can be efficiently managed \cite{9567867,abbas22,other_circuit}. The next paragraphs present an overview of the literature on CEE and GRAND decoders.

\subsection{Pilots, fading channel estimation and CEE}
Channel estimation has long been a known topic of wireless communication research, with a variety of algorithms developed, ranging from traditional methods like least-square error and maximum a posteriori approaches \cite{Heath_lecture_notes,4357450} to machine learning-based methods \cite{8052521} and blind estimation techniques \cite{4138046}. In pilot-based channel estimation, the receiver's channel estimate can be imperfect due to two major factors. First, additive noise from electronics affects the received sequence, as previously mentioned. Second, in large-scale networks, the same pilot sequence may be assigned to multiple access points due to the high number of nodes and limited number of different pilot sequences. While distinct pilot sequences are designed to be orthogonal, access points using the same pilot sequence interfere with each other, which reduces the accuracy of their respective channel estimates. This phenomenon, known as pilot contamination, is mathematically modeled in \cite{6798744} and \cite{8845768} for cellular and cell-free networks.

The impact of channel estimation errors has also been studied from an information-theoretic perspective, with expressions for capacity, mutual information, and outage probability derived in \cite{1193803,781400,4215137,5208548}. Various solutions have been proposed to jointly improve channel estimation and error correction, including methods using polar codes \cite{9492100,9082596} and turbo codes \cite{1532478,6919284}. General approaches to incorporate the statistical distribution of CEE in log-likelihood ratio (LLR) computations are also presented in \cite{6199941}.

\subsection{GRAND}
GRAND is a newly developed decoder, initially designed for hard decision demodulation receivers~\cite{8630851}. Unlike other channel decoders, GRAND seeks to identify the binary noise sequence impacting the transmission without depending on a specific code structure. This algorithm generates potential binary noise sequences in decreasing likelihood order and progressively checks whether the remaining sequence is a codebook member after subtracting the noise sequence from the demodulated sequence. In hard decision scenarios, the query order is based on statistical channel knowledge \cite{9766201}. In soft decision cases, soft information consisting of LLRs dictates the query order. Soft-GRAND (SGRAND) \cite{9149208} performs maximum likelihood decoding when soft inputs are available. While not well-suited for efficient hardware implementation, it allows for the empirical assessment of the optimal performance when computational factors are not considered. Ordered Reliability Bits Guessing Random Additive Noise Decoding (ORBGRAND) \cite{9872126} is a soft decision GRAND decoder that approximates rank-ordered bit reliabilities using piecewise linear functions, allowing for the efficient generation of putative noise sequences through generative integer partition algorithms. ORBGRAND is therefore computationally efficient and has been shown to nearly achieve the capacity in AWGN channels \cite{9992258}. GRAND-EDGE and ORBGRAND-EDGE (Erasure Decoding by Gaussian Elimination) were introduced in \cite{10279273} to address potential jamming effects. Hard decision GRAND and soft decision ORBGRAND are able to effectively decode any code with moderate redundancy, regardless of length, and are naturally suited for hardware implementation thanks to their high parallelizability~\cite{9567867,abbas22,other_circuit}.

\subsection{Contributions \& Notations}
The primary contributions of this work are as follows:
\begin{itemize} 
\item We propose a novel method that jointly performs error correction and mitigates the impact of CEE on BLER. On top of the initial channel estimate obtained using pilot signals, our method considers alternative candidates for the actual fading channel value. These additional candidates are strategically selected in the vicinity of the original estimate based on the CEE distribution. Using posterior likelihood expressions derived for ORBGRAND, we demonstrate that choosing the candidate with the highest posterior likelihood significantly reduces the BLER. 
\item We introduce an enhanced version of our method, which accounts for the presence of residual CEE in the computation of soft information when considering a candidate. This refined approach provides further performance gains. 
\item We offer an initial performance assessment of our methods for several coding schemes, including Cyclic Redundancy Check (CRC) code and CRC-Assisted Polar (CA-Polar) code, as specified in the 5G standard. 
\end{itemize}

In the upcoming sections, $\mathbb{F}_m$ denotes the Galois Field containing $m$ elements, while $\vec{x}$ and $x_i$ refer to a complex vector and its $i$th component, respectively. 

\section{System model and Background}
\label{sect:System model and Detectors}

\subsection{System model}
We consider a point-to-point communication system represented in equivalent baseband. At the transmitter, a binary information word $\vec{u} \in \mathbb{F}^k_2$ is encoded using a specified error-correcting code $\xi : \mathbb{F}^k_2 \rightarrow \mathbb{F}^n_2$ with $n > k$. The resulting codeword, denoted by $\vec{c} \in \mathbb{F}^n_2$, is part of the codebook $\mathcal{C}$, which includes all potential outputs of the encoder $\xi$. Thus, the codebook is defined as $\mathcal{C} = {\vec{c} : \vec{c} = \xi(\vec{u}), \vec{u} \in \mathbb{F}^k_2}$. Prior to analog transmission, constellation mapping is applied: each group of $m$ consecutive bits in $\vec{u}$ is mapped to a complex symbol within a constellation $\mathcal{X}$. The complex vector generated after mapping is represented by $\vec{x} \in \mathbb{C}^{n/m}$, with $m$ assumed to divide $n$. After transmission, the signal obtained at the receiver side is given by $\vec{y} = h\vec{x} + \vec{w} \in \mathbb{C}^{n/m}$. In this last equation, the following notations have been introduced:
\begin{itemize}
    \item $\vec{w}$ is a random vector representing complex additive white Gaussian noise (AWGN) of variance $\sigma_w^2$.
    \item $h$ represents the effect of the small-scale fading channel. Assuming flat fading, this coefficient remains constant during the transmission of each codeword. In terms of statistical distribution, we consider Rice fading, with Rayleigh as a particular case. The complex random variable $h$ is also normalized such that $\mathbb{E}\big[|h|^2\big]=1$, accounting for the fact that, on average, fading does not increase the energy of the transmitted symbols.
    We denote by $\hat{h}$ and $h_E$, the estimate of $h$ available at the receiver and the associated CEE. One has by definition $h = \hat{h}+h_E$. As in most existing studies \cite{6199941}, we consider $h_E$ modeled as a complex Gaussian with variance  $\sigma_E^2$.
\end{itemize}
\subsection{Detectors}
Following the same line of thought as in \cite{10001707}, we consider three different detectors applicable on the symbol vector $\vec{y}$ obtained at the receiver:
\noindent

\subsubsection{Maximum likelihood (ML)} this detector performs an exhaustive search to find the vector $\vec{x}^{\text{ML}}$ whose entries belong to the constellation $\mathcal{X}$ and which satisfies the following minimum distance criterion:
\begin{equation}
    \vec{x}^{\text{ML}} = \text{argmin}_{\vec{v} \in \hat{\mathcal{X}}} |\vec{y}-\hat{h}\vec{v}|^2.
\end{equation}
In this equation, $\hat{\mathcal{X}}$ is the set of all vectors of size $n/m$ whose entries belong to $\mathcal{X}$. Using this detector, the log-likelihood ratio (LLR) associated with the $j$th bit of the $i$th symbol is given by \small
\begin{align}
    &\lambda_{i,j}^{\text{ML}}\mathord{=}\log \dfrac{\text{Pr}\big[c_{i,j}\mathord{=}1|y_i,\!\hat{h}\big]}{\text{Pr}\big[c_{i,j}\mathord{=}0|y_i,\!\hat{h}\big]}\mathord{=}\log\frac{\sum\limits_{v_i \in \mathcal{X}^{j,1}}\!\!e^{-\frac{1}{\sigma_w^{2}}|y_i-\hat{h}v_i|^2}}{\sum\limits_{v_i \in \mathcal{X}^{j,0}}\!\!e^{-\frac{1}{\sigma_w^{2}}|y_i-\hat{h}v_i|^2}},
    \label{original_llr_ml}
\end{align}
\normalsize
where constellation symbols assumed to be equiprobable. In this last equality, $\mathcal{X}^{j,1}$ and $\mathcal{X}^{j,0}$ represent the sets of symbols within $\mathcal{X}$ for which the $j$th bit is 1 and 0, respectively. Thanks to the Jacobian-logarithm approximation $\log(\sum_r a_r) \approx \max_r \{a_r\}$, the LLR expression of \eqref{original_llr_ml} can be approximated as \cite{7436797,8186206}
\begin{equation}
    \lambda_{i,j}^{\text{ML}} \approx \dfrac{1}{\sigma_w^2} \Big( \min\limits_{v_i \in \mathcal{X}^{j,1}} |y_i-\hat{h}v_i|^2 - \min\limits_{v_i \in \mathcal{X}^{j,0}} |y_i-\hat{h}v_i|^2\Big).
    \label{eq:ML_simplified}
\end{equation}
\subsubsection{Zero forcing (ZF)}this linear detector equalizes the channel by multiplying it by the the inverse of its estimate $\hat{h}^{-1}$. The output of this equalizer is therefore given by
\begin{equation}
    y_i^{\text{ZF}} = \hat{h}^{-1} y_i = x + \hat{h}^{-1}h_{E} x_i + \hat{h}^{-1} n_i
    \label{eq:ZF_Decoder}
\end{equation}
Following the same reasoning as the ML case, the LLR values obtained with this detector can be expressed as
\begin{equation}
    \lambda_{i,j}^{\text{ZF}} \approx \dfrac{1}{\sigma_w^2/|\hat{h}|^2} \Big( \min\limits_{v_i \in \mathcal{X}^{j,1}} |y^{\text{ZF}}_i-v_i|^2 - \min\limits_{v_i \in \mathcal{X}^{j,0}} |y^{\text{ZF}}_i-v_i|^2\Big)
    \label{eq:ZF_simplified}
\end{equation}
\subsubsection{Minimum mean square error (MMSE)} this linear detector produces the following output:
\begin{equation}
    y_i^{\text{MMSE}} = (\hat{h}^*\hat{h}+\sigma_w^2)^{-1}\hat{h}^* y_i.
\end{equation}
The corresponding LLR values are given by
\begin{align}
    \lambda_{i,j}^{\text{MMSE}} \approx& \dfrac{1}{\sigma_w^2(|\hat{h}|^2+\sigma_w^2)^{-1}}\nonumber \\ 
    &\times \Big( \min\limits_{v_i \in \mathcal{X}^{j,1}} |y^{\text{MMSE}}_i\mathord{-}v_i|^2 \mathord{-} \min\limits_{v_i \in \mathcal{X}^{j,0}} |y^{\text{MMSE}}_i\mathord{-}v_i|^2\Big)
    \label{eq:MMSE_simplified}
\end{align}
\subsection{ORBGRAND}
ORBGRAND utilizes reliability values, defined as the absolute values of LLRs computed above, as soft information. Based on these values, ORBGRAND ranks the bits of the received information block from least to most reliable. It then generates potential noise sequences and subtracts them from the received codeword until a codebook member is found. The order in which these noise sequences are generated is optimal if the ordered reliability values increase linearly. This condition is shown in \cite{9872126} to hold fairly accurately under AWGN conditions at low signal-to-noise ratios. In the same reference, analytical expressions for the a posteriori likelihood of the guesses are provided. Let $\vec{z}$ denote the first tested binary noise sequence for which a codebook member is found. Redefining the notation from the previous section, let $\lambda_l$ represent the LLR of the $l$th bit of the codeword (with $l = 1, \hdots, n$). The posterior likelihood of $\vec{z}$ can be expressed as
\begin{equation}
    \mathbb{P}(\vec{Z} = \vec{z}) = \prod_{l:z_{l}=0} (1-B_{l}) \prod_{l:z_{l}=1} B_{l}
    \label{posterior_formula}
\end{equation}
with
\begin{equation}
B_{l} = \dfrac{e^{-|\lambda_l|}}{1+e^{-|\lambda_l|}}.
\end{equation}
\section{Proposed decoding methods}
\label{sect:Proposed decoding methods}

\subsection{Method 1 - No residual CEE assumed}
\begin{figure}[h!]
    \centering
    \includegraphics[width = 0.40\textwidth]{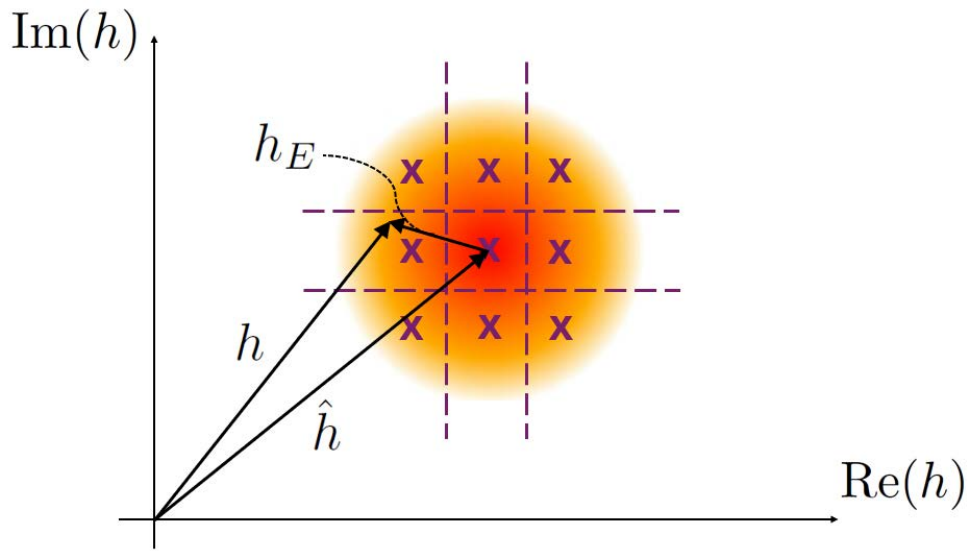}
    \caption{Illustration of the fading channel along with nine candidate estimates in the complex plane.}
    \label{fig:illustration_candidates}
\end{figure}

As mentioned above, a theoretical expression for the ORBGRAND decoder exists to quantify confidence in a guessed codeword. This first method relies on that expression, assuming that the CEE distribution is known at the receiver. Our approach involves using alternative values for the fading channel estimate when computing the LLRs necessary for decoding. These alternative values, referred to as channel candidates, are strategically positioned in the complex plane around the original pilot-based estimate, $\hat{h}$. Their placement depends on the CEE distribution and the number of candidates chosen. An example of this approach is illustrated in Fig. \ref{fig:illustration_candidates}, showing an initial estimate in the first quadrant of the complex plane, with eight additional candidates. For a total of $M$ candidates (including the initial estimate), we define $\{\hat{h}^{(m)}\}_{m=1}^{M}$ as the set of channel candidates and $\{\Delta_C^{(m)}\}_{m=1}^{M}$ as the set of differences between each candidate and the original estimate. By definition, each candidate estimate is given by $\hat{h}^{(m)} = \hat{h} + \Delta_C^{(m)}$. In practice, the set $\{\Delta_C^{(m)}\}_{m=1}^{M}$ is predetermined based on the known CEE distribution, and the candidate values $\{\hat{h}^{(m)}\}_{m=1}^{M}$ are computed and updated according to the current pilot-based estimate. For each $\hat{h}^{(m)}$, a unique vector of LLR values is obtained, distinct from those of other candidates, using either \eqref{eq:ML_simplified}, \eqref{eq:ZF_simplified}, or \eqref{eq:MMSE_simplified}. Consequently, the guessed codewords derived by ORBGRAND for each candidate may vary. In addition, the associated posterior probability values also differ. This method selects the codeword (and its corresponding channel candidate) associated with the highest posterior probability. In other words, we choose the candidate that would result in the most probable bit sequence if selected. This approach is logical, as candidates farther from the true fading value result in larger CEEs. As shown in \eqref{eq:ZF_Decoder}, these CEEs introduce an additional equivalent noise term to the pure signal, on top of the original electronic noise. Consequently, large CEEs can increase the likelihood of bit flips, corresponding to putative noise sequences that are less probable. Our method is summarized in Algorithm 1 below and numerically demonstrated in Section \ref{sect:numerical results}.

\begin{algorithm}
\caption{First Method (no residual CEE assumed)}\label{alg:cap}
\begin{algorithmic}
\Require $\vec{y}$, $\sigma_w$, $\sigma_E$, $\{\hat{h}^{(m)}\}_{m=1}^{M} \; \; \text{(the set of channel candidates)}$ 
\For{$m=1:M$}

\State{- Compute vector of LLR values $\vec{\lambda}^{(m)}$ by using  $\hat{h}^{(m)}$} 
\State{\hspace{0.13 cm} and $\vec{y}$ in  \eqref{eq:ML_simplified}, \eqref{eq:ZF_simplified} or \eqref{eq:MMSE_simplified}.}
\State{- Using $\vec{\lambda}^{(m)}$ as soft input, apply ORBGRAND on $\vec{y}$ to} \State{\hspace{0.25 cm}produce a decoded codeword $\vec{q}^{\;(m)}$ and its associated} \State{\hspace{0.25 cm}noise sequence $\vec{z}^{\;(m)}$.}
\State{- Compute the posterior likelihood $P^{(m)}$ by injecting} 
\State{\hspace{0.25 cm}$\vec{\lambda}^{(m)}$ and $\vec{z}^{\;(m)}$ in \eqref{posterior_formula}.}
\EndFor
\State - Compute $m^* = \text{argmax}_{m} \; P^{\;(m)}$
\Ensure $\vec{q}^{\;(m^*)}$, $\hat{h}^{(m^*)} $
\end{algorithmic}
\end{algorithm}

\begin{remark}
As a reminder, the CEE distribution of the system model is isotropic Gaussian. Consequently, channel candidates further from the center of the two-dimensional probability density function (pdf) are statistically less probable. To further incorporate this aspect into the channel candidate selection, it is possible to adjust the weights associated with these candidates. In the above algrithm, these weights were denoted by the variables $P^{(m)}$ and calculated from \eqref{posterior_formula}. An alternative expression is given by $P^{(m)}.f_{h_E}(\Delta_C^{(m)})$, where $f_{h_E}(\cdot)$ represents the Gaussian pdf of the CEE. This choice of weighting corresponds to the hypothesis $\mathbb{P}\big[\vec{z}^{\;(m)},\Delta_C^{(m)}\big] = \mathbb{P}\big[\vec{z}^{\;(m)}\big] \mathbb{P}\big[\Delta_C^{(m)}\big]$. As shown in Section \ref{sect:numerical results}, this weighting adjustment can yield additional performance improvements.
\label{changing_weight}
\end{remark}
\subsection{Method 2 - Residual CEE assumed}
Unlike Method 1, the approach used here does not assume that the actual fading channel value is exactly equal to the tested channel candidate when computing LLRs. Instead, it is assumed in each parallel thread that the true channel value lies within the Voronoï cell of the complex plane associated with the tested candidate. These Voronoï cells, delimited by the dashed lines in Fig. \ref{fig:illustration_candidates}, partition the complex plane. Under the Gaussian assumption of the initial CEE as described in Section \ref{sect:System model and Detectors}, the residual CEE associated to each candidate also follows a Gaussian pdf, but truncated according to the 2D boundaries of its respective Voronoï cell. For maximum-likelihood (ML) detection, the residual CEE statistics can be incorporated into the LLR computation using the approach of \cite{6199941}. This method involves integrating the conditional probabilities in the LLR definition of \eqref{original_llr_ml} over the distribution of the residual CEE. For the $m$th tested candidate, \eqref{original_llr_ml} thus becomes:

{\footnotesize
\begin{align}
    \lambda_{i,j}^{\text{ML},(m)}&\mathord{=}\log \dfrac{\int\text{Pr}\big[c_{i,j}\mathord{=}1|y_i,\!\hat{h}^{(m)}\mathord{+}\Delta h^{(m)}_E\big] f\big(\Delta h^{(m)}_E\big) d\Delta h^{(m)}_E}{\int\text{Pr}\big[c_{i,j}\mathord{=}0|y_i,\!\hat{h}^{(m)}\mathord{+}\Delta h^{(m)}_E\big]f\big(\Delta h^{(m)}_E\big) d\Delta h^{(m)}_{E}} 
\label{eq_method_2}
\end{align}}

\normalsize
where $\Delta h_{E,m}$ represents the residual CEE for the $m$th candidate, and $f(\Delta h_{E,m})$ is its probability density function. The integrals in \eqref{eq_method_2} are two-dimensional over the complex plane. The limits of these integrals, along with the analytical expressions obtained from calculating them, depend on the boundaries of the Voronoï cell. Apart from this modification in the LLR computation, the rest of Algorithm 1 remains unchanged.

\section{Numerical results}
\label{sect:numerical results}
\begin{figure}[h!]
    \centering
    \includegraphics[width = 0.48\textwidth]{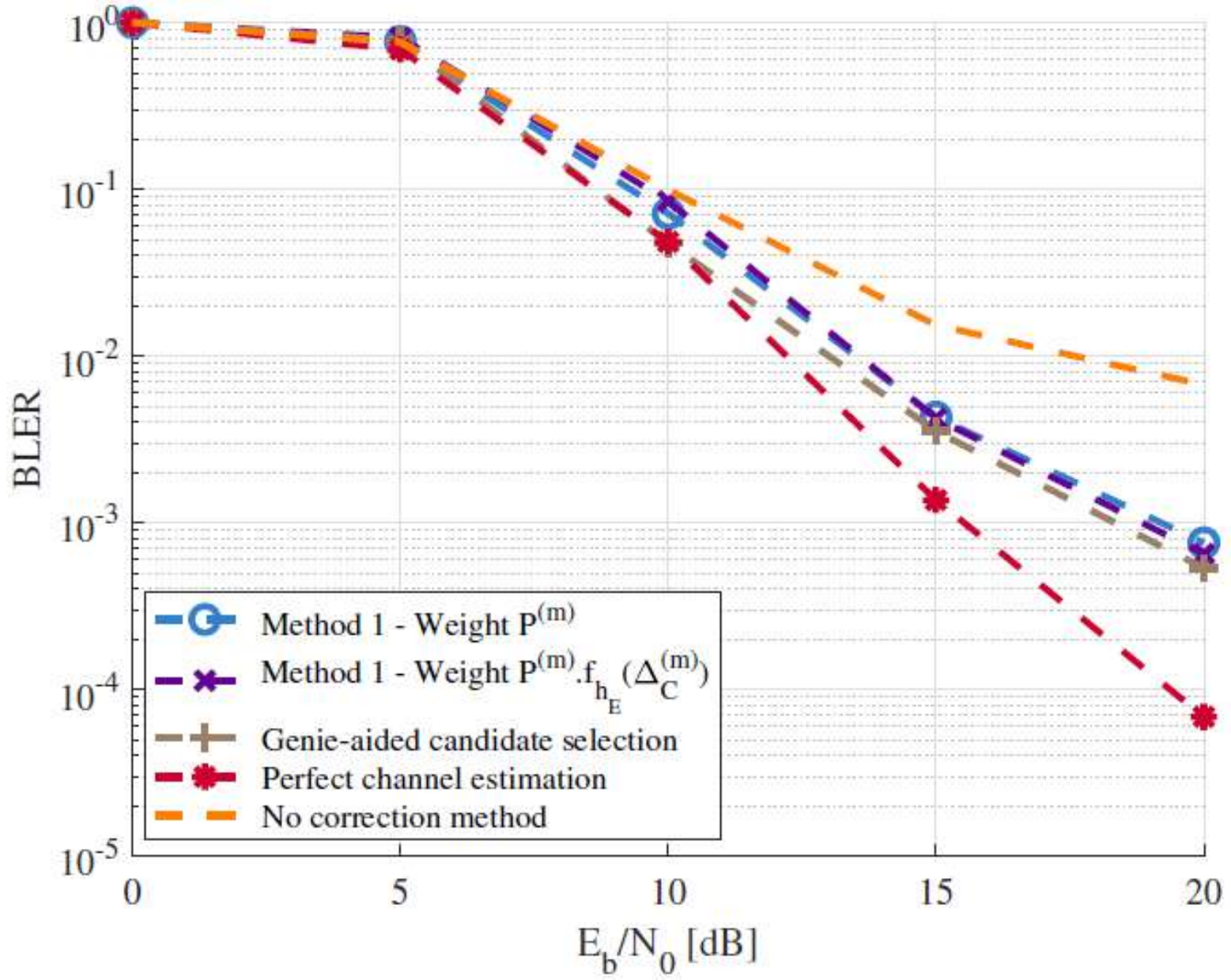}
    \caption{Performance of Method 1 for a [128,112] CRC code.}
    \label{fig:perf_method_1_CRC}
\end{figure}

\begin{figure}[h!]
    \centering
    \includegraphics[width = 0.48\textwidth]{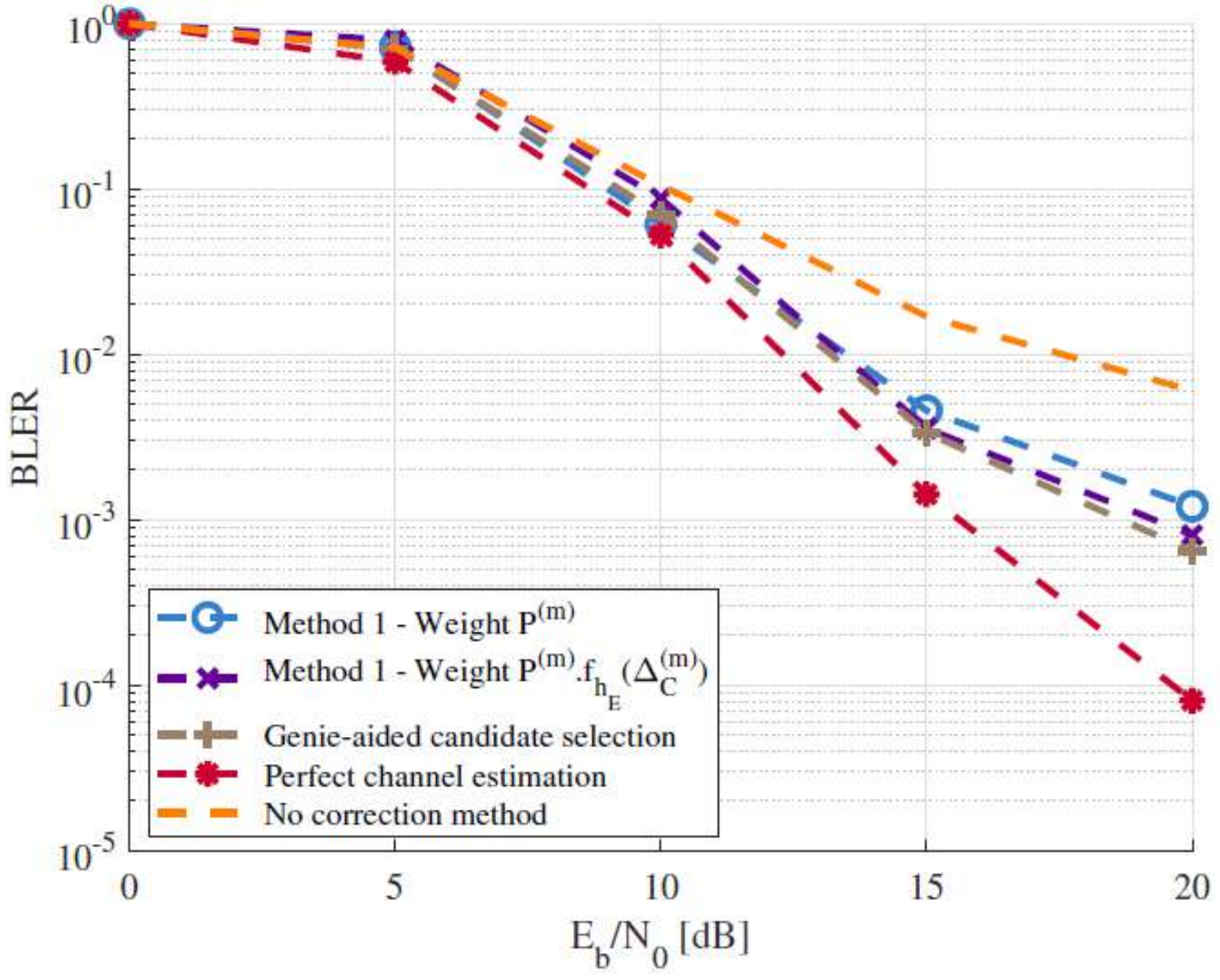}
    \caption{Performance of Method 1 for a [128,112] CA-Polar code.}
    \label{fig:perf_method_1_CAPOLAR}
\end{figure}

The performance of Method 1 is depicted in Figs. \ref{fig:perf_method_1_CRC} and \ref{fig:perf_method_1_CAPOLAR}. These graphs have been generated using two families of codes that have demonstrated competitive results with GRAND algorithms \cite{9500279,9086275}. Fig. \ref{fig:perf_method_1_CRC} presents results obtained using a CRC code, traditionally applied for error detection. The motivation for its use in error correction and its performance when employed with GRAND is discussed in \cite{9500279}. Figure \ref{fig:perf_method_1_CAPOLAR} illustrates results with a CRC-assisted polar (CA-Polar) code. This family of codes is commonly used for control channel communications in 5G New Radio and has also been analyzed with GRAND decoders \cite{9086275}. To ensure consistency, both figures use code dimensions $[n,k] = [128,112]$. The CRC length of the the CA-Polar code is given by $\ell = 11$. For the other parameters, we considered $\sigma_E^2 = 0.01$, Rice fading with a K-factor of 10, a MMSE detector, and a 16-QAM constellation. Five candidate channels were considered, corresponding to
\begin{equation}
    \big\{\Delta_C^{(m)}\big\}_{m=1}^{M} = \bigg\{0, \dfrac{\sigma_E}{\sqrt{2}},-\dfrac{\sigma_E}{\sqrt{2}},\dfrac{\sigma_E}{\sqrt{2}}j,-\dfrac{\sigma_E}{\sqrt{2}}j \bigg\}.
\end{equation}

In both figures, a gain of over 5 dB is visible in the 15-20 dB range when comparing the performance of Method 1 (blue curve) with the BLER obtained without any correction for the CEE (orange curve). As noted in Remark \ref{changing_weight}, the error rate slightly decreases when refining the selection weights to further account for the CEE distribution. It can also be observed that the results of Method 1 closely approach the genie-aided performance shown in grey. This benchmark is achieved by selecting the candidate that minimizes the number of bit errors between its associated codeword $\vec{q}^{\;(m)}$ and the original codeword $\vec{c}$ sent by the transmitter. These results are therefore genie-aided as they would require the prior knowledge of $\vec{c}$ at the receiver. 

\begin{figure}[h!]
    \centering
    \includegraphics[width = 0.48\textwidth]{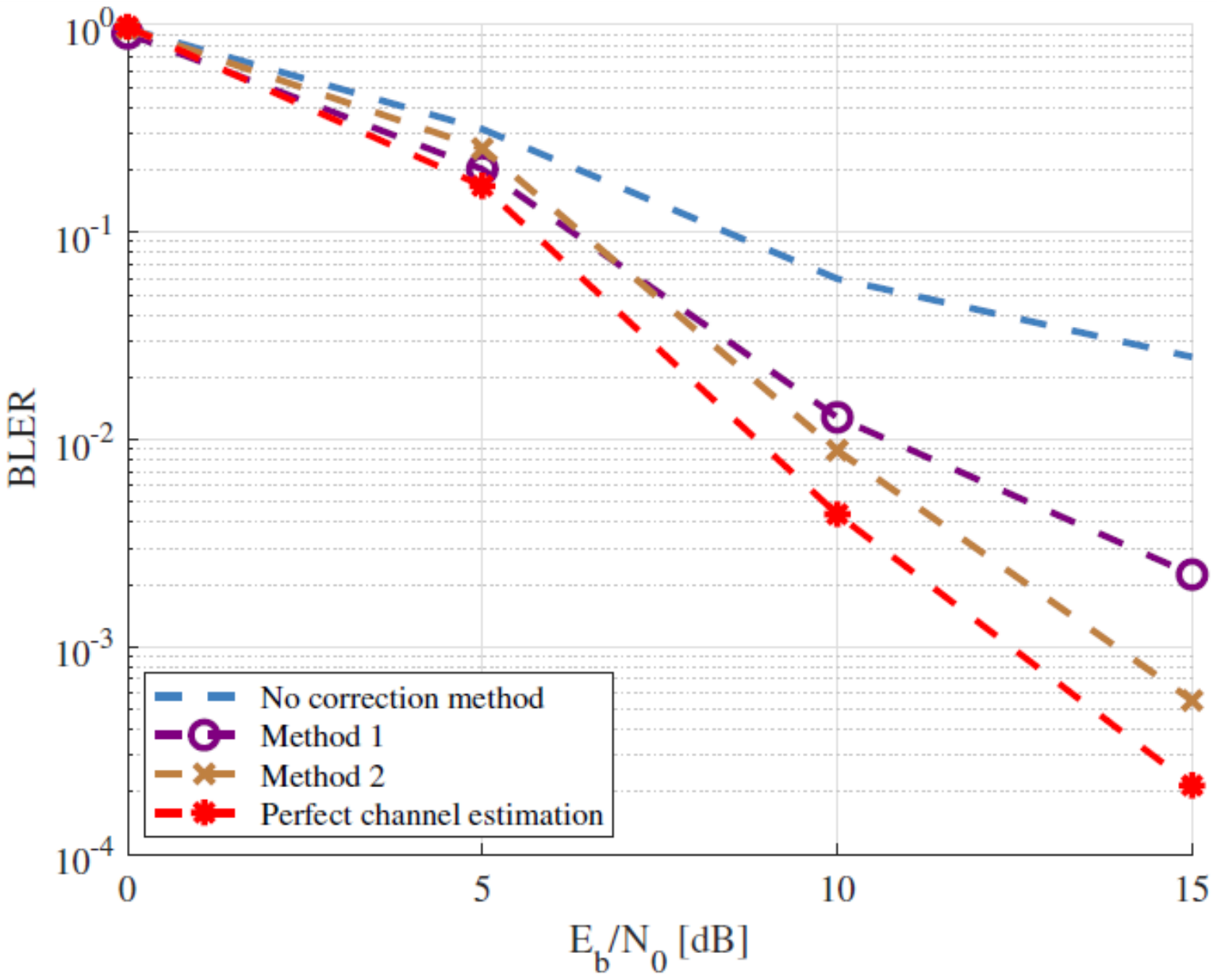}
    \caption{Performance of Method 2 for a [128,112] CRC code.}
    \label{fig:perf_method_2}
\end{figure}

The BLER associated with Method 2 is shown in Fig. \ref{fig:perf_method_2}. This figure is generated with parameters $\sigma_E^2 = 0.1$, Rice fading with a K-factor of 10, an ML detector, a QPSK constellation, and a [128,112] CRC code. The nine candidate channel values illustrated in Fig. \ref{fig:illustration_candidates} were considered, spaced horizontally and vertically by $\sigma_E/\sqrt{2}$. This choice of candidates simplifies the Voronoï cell structure, allowing closed-form expressions for the integrals in \eqref{eq_method_2} in terms of the error function. With these settings, Method 2 achieves an additional BLER gain of approximately 2 dB compared to Method 1.

\section{Conclusion}
In this study, we introduced new approaches to improve error correction in scenarios with channel estimation errors. These methods were tested across several code families, demonstrating substantial performance gains—amounting to several dB in terms of BLER. The parallelizable nature of our approach enables wireless transceivers to achieve the same performance with less precise fading estimations. This capability would allow for shorter pilot sequences in channel estimation, thereby enhancing data rates. Consequently, this method introduces new trade-offs between data rate and system complexity.

In future work, we aim to explore additional advantages of our approach in time-varying channels with temporal correlation. Our approaches could potentially track channel evolution and adapt over time to provide increasingly accurate estimations. We also plan to enhance our second method by leveraging the received constellation symbols. Instead of defaulting to an isotropic Gaussian pdf, a more accurate conditional distribution for the CEE could be computed based on these symbols.

\section*{Acknowledgment}
This work was supported by the Defense Advanced Research Projects Agency (DARPA) under Grant HR00112120008. Charles Wiame was supported by a Fellowship of the Belgian American Educational Foundation. 

\bibliographystyle{IEEEtran}
\bibliography{bibli}

\end{document}